\def\draft{0}

\documentclass[11pt]{article}
\usepackage{amsfonts, amsmath, amssymb, fullpage, url}

\newtheorem{theorem}{Theorem}[section]

\newtheorem{definition}[theorem]{Definition}
\newtheorem{lemma}[theorem]{Lemma}
\newtheorem{proposition}[theorem]{Proposition}

\newtheorem{remk}[theorem]{Remark}

\def\FullBox{\hbox{\vrule width 8pt height 8pt depth 0pt}}

\def\qed{\ifmmode\qquad\FullBox\else{\unskip\nobreak\hfil
\penalty50\hskip1em\null\nobreak\hfil\FullBox
\parfillskip=0pt\finalhyphendemerits=0\endgraf}\fi}

\newenvironment{proof}{\begin{trivlist} \item {\bf Proof:~~}}
  {\qed\end{trivlist}}

\def\qedsketch{\ifmmode\Box\else{\unskip\nobreak\hfil
\penalty50\hskip1em\null\nobreak\hfil$\Box$
\parfillskip=0pt\finalhyphendemerits=0\endgraf}\fi}

\ifnum\draft=1
\newcommand{\authnote}[2]{{ \bf [#1's Note: #2]}}
\else
\newcommand{\authnote}[2]{}
\fi

\newcommand{\COMMENT}[1]{}
\newcommand{\ket}[1]{|#1\rangle}
\newcommand{\bra}[1]{\langle#1|}
\newcommand{\ketbra}[2]{|#1\rangle\langle#2|}
\def\01{\{0,1\}}
\def\01{\{0,1\}}
\newcommand{\prot}[1]{\langle {#1} \rangle}

\newcommand{\cadre}[1]
{
\begin{tabular}{|p{15.4cm}|}
\hline
#1 \\
\hline
\end{tabular}
}

\newcommand{\spa}[1]{\mathcal{#1}}

\newcommand{\modif}[1]{#1}
 
\begin{document}
\title{Increasing the power of the verifier in Quantum Zero Knowledge}

% These are required for headers and for copyright on title page
%\runningtitle{Increasing the power of the verifier in Quantum Zero Knowledge}
%\runningauthors{Chailloux, Kerenidis}

\author{Andr\'e Chailloux$^*$ \\
LRI\\
Universit\'e Paris Sud \\
andre.chailloux@lri.fr\\
\and
Iordanis Kerenidis\thanks{Supported in part by ACI Securit\'e Informatique SI/03 511 and ANR AlgoQP grants of the French Ministry and in part by the European Commission under the Integrated Project Qubit Applications (QAP) funded by the IST directorate as Contract Number 015848.}\\
CNRS - LRI\\
Universit\'e Paris-Sud\\
jkeren@lri.fr
}

%\affiliation{
%  CNRS - LRI\\
%  Universit\'e Paris-Sud\\
%  \email{\{chaillou,jkeren\}@lri.fr}
%} 

\maketitle
\thispagestyle{empty}

%\tableofcontents

\begin{abstract}

In quantum zero knowledge, the assumption was made that the verifier is only using unitary operations. Under this assumption, many nice properties have been shown about quantum zero knowledge, including the fact that Honest-Verifier Quantum Statistical Zero Knowledge ($HVQSZK$) is equal to Cheating-Verifier Quantum Statistical Zero Knowledge ($QSZK$) (see ~\cite{Wat02,Wat06}).

In this paper, we study what happens when we allow an honest verifier to flip some coins in addition to using unitary operations. Flipping a coin is a non-unitary operation but doesn't seem at first to enhance the cheating possibilities of the verifier since a classical honest verifier can flip coins. In this setting, we show an unexpected result: any classical Interactive Proof has an Honest-Verifier Quantum Statistical Zero Knowledge proof with coins. Note that in the classical case, honest verifier $SZK$ is no more powerful than $SZK$ and hence it is not believed to contain even $NP$. On the other hand, in the case of cheating verifiers, we show that Quantum Statistical Zero Knowledge where the verifier applies any non-unitary operation is equal to Quantum Zero-Knowledge where the verifier uses only unitaries.

One can think of our results in two complementary ways.
If we would like to use the honest verifier model as a means to study the general model by taking advantage of their equivalence, then it is imperative to use the unitary definition without coins, since with the general one this equivalence is most probably not true.
On the other hand, if we would like to use quantum zero knowledge protocols in a cryptographic scenario where the honest-but-curious model is sufficient, then adding the unitary constraint severely decreases the power of quantum zero knowledge protocols. 

\end{abstract}

\newpage
\setcounter{page}{1}

\section{Introduction}

Zero knowledge protocols propose an elegant way of doing formally secure identification. In these interactive protocols, a prover $P$ knows a secret $s$ and he wants to convince a verifier $V$ that he knows $s$ without revealing any
information about $s$. The condition "without revealing any information" has been formalized in \cite{GMR89,GMW91} and this security condition has been defined in the computational $(CZK)$ and the information-theoretic setting ($SZK$). Zero knowledge has been extensively studied and found numerous applications in theoretical computer science and cryptography (see \cite{Vad99} and references therein).

In addition, zero knowledge is defined for the case of honest or cheating verifiers. In the honest verifier model, we force the protocol to be zero knowledge only against a verifier who follows the protocol but tries to extract as much information as possible from the interaction. An honest verifier is equivalent to the `Honest-but-Curious' or `Semi-Honest' adversary in cryptography.  This model has been widely studied in cryptography and is important in certain realistic scenarios (for example online protocols), where the protocols are used in complex interactions with limited capacity of cheating (\cite{Gol01}, ch. 7).
Moreover, in the case of classical zero knowledge it is particularly interesting, due to the fact that it is equivalent to the general Zero-Knowledge model against cheating verifiers\modif{~\cite{GSV98}}.

In $2002$, Watrous proposed a quantum equivalent of zero knowledge proofs \cite{Wat02} for the case of honest verifiers. In this definition, the prover and the verifier are allowed to use only unitary operations and the zero knowledge property is defined in a seemingly weaker way than in the classical case (see also Section 2). Watrous proved many interesting results for this class, such as complete problems, closure properties and a few years later, that honest verifier equals cheating verifier (i.e. $HVQSZK = QSZK$) \cite{Wat06}. These results provided strong {\em a posteriori} evidence that Watrous' definition is the right one for quantum Zero Knowledge.

In this paper, we revisit the definition of quantum zero knowledge and examine the importance of the unitarity constraint. First, we increase the power of the honest verifier by allowing him to flip classical coins in addition to performing unitary quantum operations. Note that flipping classical coins is not a unitary operation and that coin flips are also allowed in the classical case. In this new setting, we also strengthen the definition of simulation in order to still catch the essence of Zero-Knowledge protocols. In particular, the verifier does not "forget" or "erase" these coins, since he remains honest but curious. Even though this augmentation to the model seems minimal if not trivial, we prove that any classical interactive proof has a quantum honest-verifier statistical zero-knowledge proof (Section \ref{HVQSZK}) with coins. Note that in the classical case, honest verifier $SZK$ is no more powerful than $SZK$ and hence it is not believed to contain even $NP$. Our proofs go through the notion of "hidden-bits" which has been previously studied in \cite{FLS00} using ideas from~\cite{Kil88}.

If, on the other hand, we look at cheating verifiers, we show that the most general cheating strategies for quantum verifiers are the unitary ones. In Section \ref{QSZK_NU}, we transform any general Zero Knowledge protocol into a unitary protocol that retains completeness, soundness and the zero-knowledge property.

We like to see the consequences of our results from two different points of view. On one hand, if we want to use the honest verifier model as a means for the study of general zero knowledge, then the most important property that we would like is the equivalence of the two models. This way, one only needs to prove that a protocol is zero knowledge against honest verifiers and immediately conclude that it can also be made zero knowledge against cheating verifiers. Our results show that in this case, Watrous' definition with unitaries is indeed the right one, since we give strong evidence that this equivalence does not hold in the non-unitary case. Moreoover, we prove that the use of non-unitaries does not change the power of a cheating verifier.

On the other hand, the Honest-but-Curious model (that corresponds to the honest verifier) is not only a means for the study of the malicious model (that corresponds to the cheating verifier) but an important model in itself pertinent to many realistic cryptographic scenarios. For example, in certain settings, we can assume that the verifier is semi-honest when he interacts with the prover via a secure interface, eg. an ATM or a secure web interface. In this case, it might suffice to assume that the verifier does not open the ATM by force or hack the webpage, instead he can only provide well-chosen legal inputs to these machines and try to extract as much information as possible from the interaction.

\paragraph{Organization of our work} $ \ $
\begin{itemize}
\item In \textbf{Section \ref{prel}}, we first present the class $HVQSZK$ originally defined by Watrous where the verifier is allowed to use only unitary operations. We then extend this definition to the case where the verifier can use only unitaries and flip some classical coins, resulting in the class $HVQSZK^{C}$. Finally, we introduce the notion of hidden bits defined and used in $\cite{FLS00}$ and $\cite{PS05}$.
\item In \textbf{Section \ref{HVQSZK}}, we show the main result of our work: $PSPACE \subseteq HVQSZK^C$, and discuss the applicability of the semi-honest model.
\item In \textbf{Section \ref{QSZK_NU}}, we show that in the context of cheating verifiers, allowing the verifier to apply non-unitaries is no more powerful than allowing him to use only unitaries. \modif{In other words}, we show that $QSZK^{U} = QSZK$.
\end{itemize}

%%%%%%%%%%%%%%%%%%%%%%%%%%%%%%%%%%%%%%%%%%%%%%%%%%%%%%%%%%%%%%%%%
%%%%%%%%%%%%%%%%%%%%%%%%%%%%%%%%%%%%%%%%%%%%%%%%%%%%%%%%%%%%%%%%%
%%%%%%%%%%%%%%%%%%%%%%%%%%%%%%%%%%%%%%%%%%%%%%%%%%%%%%%%%%%%%%%%%
%%%%%%%%%%%%%%%%%%%%%%%%%%%%%%%%%%%%%%%%%%%%%%%%%%%%%%%%%%%%%%%%%

\section{Definitions of classical and quantum Statistical Zero Knowledge}\label{prel}

An {\em
interactive proof system} for a problem
$\Pi$ is an interactive protocol between a computationally
unbounded prover $P$ and a probabilistic polynomial-time verifier
$V$ that satisfies the following two properties:
\begin{itemize}
\item {\em Completeness:} if $x$ is a YES instance of $\Pi$ ($x \in \Pi_Y$),
then $V$ will accept with probability greater than $2/3$ after interacting
with $P$ on common input $x$.

\item {\em Soundness:} if $x$ is a NO instance of $\Pi$ ($x \in \Pi_N$), then
for every (even computationally unbounded) prover strategy $P^*$,
$V$ will accept with probability less than $1/3$ after interacting with $P^*$
on common input $x$.
\end{itemize}

\begin{definition}\label{solves}
We say that a protocol $\langle P,V \rangle$ solves $\Pi$ if and only if $\langle P,V \rangle$ is an interactive proof system for $\Pi$.
\end{definition}

In the classical Zero-Knowledge setting, we want the Verifier to learn
nothing from the interaction with the Prover, other than the fact that the input is a Yes instance of the problem ($x \in \Pi_{Y}$) when it is the case.
The way this is formalized is that for $x \in \Pi_{Y}$, one can simulate in probabilistic polynomial-time the Verifier's view of the protocol $view_{\langle P,V \rangle}(x)$, i.e. his private coins, the messages he received from the Prover and the messages he sent to the Prover. Note that the view is a distribution depending on the random coins of the Prover and the Verifier and contains all the information that the Verifier gains by interacting with the Prover. Specifically,

\begin{definition}\label{ZK property}
A protocol $\langle P,V \rangle$ has the zero-knowledge property for
$\Pi$ if there exists a probabilistic polynomial-time simulator $S$ and a negligible function $\mu$ such that for $\forall x \in \Pi_{Y}$, the simulator outputs a distribution $S(x)$ such that
$ | view_{\langle P,V \rangle}(x) - S(x) |_1 \leq {\mu(|x|)}$.
\end{definition}

In our discussion so far, we have considered the case where the Verifier honestly follows the protocol but tries to extract as much information as possible from the interaction with the Prover. In order to do that, the Honest Verifier would keep a copy of all the messages and his coins throughout the protocol and would not erase or discard any of this information.

We can now define the class of Honest Verifier Statistical Zero Knowledge ($HVSZK$):

\begin{definition}
$\Pi \in HVSZK$ iff there exists an interactive protocol $\langle P,V
\rangle$ that solves $\Pi$ and that has the zero-knowledge property
for $\Pi$.
\end{definition}

\subsection{Honest Verifier Quantum Statistical Zero Knowledge}

Quantum Statistical Zero Knowledge proofs are a special case of
Quantum Interactive Proofs. They were defined for honest verifiers by Watrous in \cite{Wat02} and have been also studied in $\cite{Kob03,Wat06,Kob07}$. We can think of a quantum interactive
protocol $\langle P,V \rangle(x)$ for a promise problem $\Pi$ as a circuit
$\left(V_{1}(x),P_{1}(x),\dots,V_{k}(x),P_{k}(x)\right)$ acting on
$\mathcal{V} \otimes \mathcal{M} \otimes \mathcal{P}$. $\mathcal{V}$
are the Verifier's private qubits, $\mathcal{M}$ the message
qubits and $\mathcal{P}$ the Prover's private qubits. $V_{i}(x)$
(resp. $P_{i}(x)$)  represents the $i^{th}$ action of the Verifier
(resp. of the prover) during the protocol and is decribed by a super-operator acting on $\mathcal{V}
\otimes \mathcal{M}$ (resp. on $\mathcal{M} \otimes \mathcal{P}$). 
$\beta_{i}$ corresponds to the state in $\mathcal{V} \otimes \mathcal{M} \otimes \mathcal{P}$ after the $i^{th}$
action of the protocol. In other words, $\beta_0$ is the initial state, $\beta_{2i}$ is the state after $P_i$ and $\beta_{2i - 1}$ the one after $V_i$.

Defining the Zero-Knowledge property in the quantum setting is not straightforward, even for the Honest Verifier case. We would still like to say that a quantum protocol has the zero knowledge property if there exists an efficient way to simulate the Verifier's view of the protocol. The main difficulty, however, is the definition of the view of the Verifier, since in the quantum case there is no notion of transcript. Indeed, the Verifier and Prover send the same qubits back and forth during the protocol and hence an Honest-but-curious Verifier cannot follow the protocol and simultaneously keep a copy of all the quantum messages that have been previously sent.

Watrous (\cite{Wat02}) tried to resolve these problems by defining honest verifier quantum zero knowledge in the following way: the view of the Honest Verifier for every round $j$ is the Verifier's part of the state $\beta_j$, i.e. $view_{\langle P,V \rangle}(j) = Tr_{\mathcal{P}}(\beta_{j})$.
We say that the Verifier's view can be simulated if there is a negligible function $\mu$ such that on any input $x$ and for each step $j$ we can
create in quantum polynomial-time a state $\sigma_{j}$ such that
$
\| \sigma_{j} - view_{\langle V,P \rangle}(j) \| \le \mu(|x|)
$.

We also distinguish the
Verifier's view depending on whether the last action was made by the Verifier or the Prover. We note $\rho_{0}$ the input state, $\rho_{i}$ the
Verifier's view after $P_{i}$ and $\xi_{i}$ the
Verifier's view after $V_{i}$.
Note that for a state $\sigma$ with $\| \sigma - \rho_{i} \|
\le \mu(|x|)$ it is easy to see that \modif{$\sigma' = V_{i+1}(\sigma)$ is close to $\xi_{i+1} = V_{i+1}(\rho_{i})$} in the sense that $\| \sigma' - \xi_{i+1} \| \le
\mu(|x|)$. Hence, we just need to simulate the $\rho_{i}$'s and hence

\begin{definition}
A protocol $\langle P,V \rangle$ has the zero-knowledge property for
$\Pi$ if there is a negligible
function $\mu$ such that $\forall x \in \Pi_{Y}$ and $\forall j$ we can create $\sigma_{j}$ with
quantum polynomial computational power such that
$
\| \sigma_{j} - \rho_{j} \|_{tr} \le \mu(|x|)
$.
\end{definition}

Let us look more closely to the `round-by-round' definition of the simulation. First, the fact that we simulate the verifier's view at every round and not just  at the end of the protocol ensures that the zero knowledge property is retained even if the Honest Verifier follows the protocol up to some round and then decides to abort.

Second, in order for this definition to be pertinent in the honest but curious model, we need to ensure that the verifier will retain all the information that he acquires during the protocol and not forget any of it. One way to ensure this is by restricting the verifier to use only unitary operations. The intuition is that since unitary operations are reversible, they do not allow for `forgetting' any information. This is precisely the way Watrous defined the class of Honest Verifier Quantum Statistical Zero Knowledge ($HVQSZK$):

\begin{definition}
$\Pi \in HVQSZK$ iff there exists a quantum protocol $\langle P,V
\rangle$ with $V$ using only unitaries that solves $\Pi$ and that has the zero-knowledge property for $\Pi$.
\end{definition}

The above intuition was later confirmed by the fact that indeed Honest Verifier Quantum Statistical Zero Knowledge with unitaries is equivalent to general cheating verifiers (\cite{Wat06}).

\subsection{The coin model for Honest Verifier Quantum Zero-Knowledge}

As we said, we would like to investigate the importance of the unitarity constraint in the power of quantum zero knowledge. For this, we define and study a new model for quantum zero-knowledge protocols, where we just allow the verifier to flip classical coins in addition to performing unitary operations. This is equivalent to saying that the verifier starts with a private random string $r^{*}$ or in a quantum language that the verifier starts with some private qubits initialized to $\ket{0}$ $-$ acting as the verifier usual workspace, and additionally some qubits in the totally mixed state $\mathbb{I}$ $-$ acting as the verifier's initial coins. The verifier uses his coins (the state $\mathbb{I}$) only as control bits. 
More formally, if we suppose that the verifier starts with the state $\mathbb{I} \otimes \ketbra{0}{0}$ in the space $\mathcal{A} \otimes \mathcal{B}$, then he can only use the space $\mathcal{A}$ by applying  unitaries of the form:
\[
U(\ket{x},\ket{y}) = \ket{x} \otimes \ket{y \oplus f(x)} \quad \textrm{with } \ket{x} \in \mathcal{A} \textrm{ and } \ket{y} \in \mathcal{B}
\]
Note that this constraint just implies that the verifier doesn't forget his coins. In particular, he does not discard these bits by sending them to the prover. 

In this case, of course, one needs to be very careful with the definition of the simulation since now, the Verifier has the extra classical information of the coins. Since the interaction is quantum we still have to consider a `round-by-round' simulation. However, in our definition of the `round-by-round' simulation we need to insist that one must simulate the entire private random string of the verifier in addition to the quantum view of the Verifier.

Note that apart from these additional initial coins, the verifier is allowed to use only unitaries like in the original definition of $HVQSZK$.
We can now define $HVQSZK^C$:

\begin{definition}
$\Pi \in HVQSZK^{C}$ iff, there exists a quantum protocol $\langle
P,V \rangle$, where the verifier's initial state is
$(\ket{0}\bra{0})^{\otimes n} \otimes \mathbb{I}_n$, that solves $\Pi$ and has the zero-knowledge property for $\Pi$. The verifier uses only unitaries and uses his coins (the state $\mathbb{I}_n$) only as control bits.
\end{definition}

\COMMENT{
\begin{definition}
$\Pi \in HVQSZK^{C}$ iff, there exists a quantum protocol $\langle
P,V \rangle$ with the restriction stated above that solves $\Pi$, that has the zero-knowledge property
for $\Pi$, where the verifier's initial state is
$(\ket{0}\bra{0})^{\otimes n} \otimes \mathbb{I}_n$. The verifier keeps at each round all his coins (all the state $\mathbb{I}_n$).
\end{definition}
}

This model is meant to be a very small augmentation of the original model proposed by Watrous. Note that the verifier is not able to create by himself the totally mixed state using only unitaries. It is important to notice that the requirement ``the prover uses the state $\mathbb{I}_n$ as control bits''  means that these coins are always part of his view of the protocol or in other words that he never forgets his coins.

\subsection{The hidden-bits model for Statistical Zero-Knowledge} \label{hidden-bits}

The hidden-bits model was first defined for Non-Interactive Zero-Knowledge \cite{FLS00}, however, it naturally extends to the interactive case.

\begin{definition}
We say that the prover has a hidden-bit $r$ with security parameter $k$ iff:
\begin{itemize}
\item $r$ is a truly random bit known to the prover.
\item The verifier has no information about $r$.
\item The prover can reveal the value of the bit $r$ to the Verifier. If he tries to convince the Verifier that the value is $\overline{r}$ then he will be caught with probability $(1 - 2^{-k})$.
\end{itemize}
\end{definition}

\begin{definition}
$\Pi \in HVSZK^{HB}$ iff there exists a classical protocol $\langle
P,V \rangle$ that solves $\Pi$ and has the zero-knowledge property
for $\Pi$ where the prover starts with a polynomial number of hidden bits.
\end{definition}

We can also \modif{define} the associated quantum class
\begin{definition}
$\Pi \in HVQSZK^{HB}$ iff there exists a quantum protocol $\langle
P,V \rangle$ that solves $\Pi$ and has the zero-knowledge property
for $\Pi$ where the prover starts with a polynomial number of hidden bits.
\end{definition}

Note that the existence of hidden-bits is a very strong assumption. In particular, we can remark that hidden-bits imply that the prover and verifier can perform bit commitment with perfect hiding and statistically binding conditions. Bit commitment is a primitive used in many cryptographic protocols. More formally:

\begin{definition} A bit commitment scheme with perfect hiding condition and statistically binding condition with security parameter $k$ is a scheme with a commit phase and a reveal phase such that:
\begin{itemize}
\item Commit phase: the prover chooses a bit $c$ and commits to it by interacting with the verifier. At the end of the interaction, the verifier has no information about $c$ (perfectly hiding).
\item Reveal phase: the prover sends a message to the verifier and reveals the commited bit $c$. If the prover tries to cheat and reveal $\overline{c}$ then he will be caught by the verifier with probability greater than $(1 - 2^{-k})$ (statistically binding).
\end{itemize}
\end{definition}

Note that both classically and quantumly, bit commitment schemes with \modif{$k \ge 1$} do not exist unconditionally\modif{~\cite{LC97,May97}}. However, there is an easy way to do bit commitment which is perfectly hiding and statistically binding with security parameter $k$ from a hidden bit $r$ with security parameter $k$. The prover commits to a bit $c$ by sending $c \oplus r$ and later reveals $r$.
After the commit phase the verifier has no information about $r$ (and hence $c$) and during the reveal phase the prover cannot lie about $r$ (and hence $c$) without being caught with probability at least $(1 - 2^{-k})$. Hence, this scheme is a commitment scheme which is perfectly hiding and statistically binding with security parameter $k$.

Classically, if we suppose the existence of such a bit commitment scheme, we can create zero-knowledge protocols for all interactive proofs \cite{BGG+90} and since Shamir showed that $IP=PSPACE$ ~\cite{Sha92}, we have
\[
PSPACE = IP \subseteq HVSZK^{HB} \]

%%%%%%%%%%%%%%%%%%%%%%%%%%%%%%%%%%%%%%%%%%%%%%%%%%%%%%%%%%%%%%%%%%%%%
%%%%%%%%%%%%%%%%%%%%%%%%%%%%%%%%%%%%%%%%%%%%%%%%%%%%%%%%%%%%%%%%%%%%%
%%%%%%%%%%%%%%%%%%%%%%%%%%%%%%%%%%%%%%%%%%%%%%%%%%%%%%%%%%%%%%%%%%%%%
%%%%%%%%%%%%%%%%%%%%%%%%%%%%%%%%%%%%%%%%%%%%%%%%%%%%%%%%%%%%%%%%%%%%%

\section{The role of coins in Quantum Statistical Zero-Knowledge}\label{HVQSZK}

In this section we start by discussing how coins have already been used in quantum zero-knowledge protocols even with the original definition of $HVQSZK$ with unitaries. We then proceed to prove our main result, that $HVSZK^{HB} \subseteq HVQSZK^C$ which implies that $PSPACE \subseteq HVQSZK^C$. Since $QMA \subseteq PSPACE$ we can already conclude that $QMA$ has an $HVQSZK$ protocol with coins. In Appendix \ref{QMA} we explicitly describe a protocol for a $QMA$ complete problem that has constant rounds and relies in fact only on the ability to do bit commitment.

\subsection{Using coins in quantum zero-knowledge protocols}\label{sofar}

Flipping a coin is a quantum non-unitary operation and therefore it should not be a priori allowed in Quantum Zero-Knowledge protocols. More precisely, a unitary quantum circuit cannot create the totally mixed state $\sum_{r \in \{0,1\}} |r\rangle\langle r|$, rather it can create the superposition of all the values of the coin as $\sum_{r \in \{0,1\}} |r\rangle$.  Note that by using this state, the verifier can still run a protocol where he is supposed to flip coins. The only difference is that his private qubits have changed and, therefore, the Zero-Knowledge property is not necessarily retained.

Coin flips have already been used in quantum Zero-Knowledge, for example in the \modif{$\overline{QSD}$,the $3$-Coloring or the Graph Isomorphism protocol} described in~\cite{Wat02, Wat06}. In these protocols, it was implied that by using only unitaries, we could create quantum states such that the protocol would work in exactly the same way as with the coin flips. This is indeed the case for these \modif{three} protocols and more generally in the following situations:
\begin{itemize}
\item \underline{The coins are made public}: in this case, the
verifier creates $\sum_{r \in \{0,1\}} |r\rangle |r\rangle$ and
sends half of this state to the prover. The prover and the verifier
both have $\sum_{r \in \{0,1\}}|r\rangle\langle r|$ which is the
desired state. This is what happens for example in the $3$-coloring
protocol in $QZK$ \modif{or in the Graph isomorphism protocol}. Note that \modif{for the 3-coloring protocol} it would not be possible to do the
simulation if the coins were in superposition.
\item \underline{The coins are not necessary for the Zero-Knowledge property}: in some protocols the Zero-Knowledge property does not depend on whether the verifier creates his coins in superposition or flips classical coins. This is what happens in the $\overline{QSD}$ protocol.
\end{itemize}

However, we will show that the situation is much more subtle than first thought. If the honest verifier is allowed coin flips, then we can show how to create hidden-bits and hence prove that $HVQSZK^{HB} \subseteq HVQSZK^{C}$. This allows us to conclude that $PSPACE \subseteq HVQSZK^{C}$.
This is a striking result that comes in contrast to the classical and the quantum unitary honest verifier $HVSZK$ and $HVQSZK$, since these classes are not known nor believed to contain even $NP$\footnote[1]{\modif{In the classical case, $NP \subseteq HVSZK$ implies that the polynomial hierarchy collapses~\cite{BHZ87} but no such results are known for the quantum case}}. The next section will be devoted to the proof of this fact.

\subsection{From coins to Hidden Bits}\label{BigProof}

We will first present a general method to create hidden-bits out of shares. We will then show a way to achieve these shares with a quantum honest verifier that has coins.

\subsubsection{A general method for creating hidden-bits}

The method described here is the one used in \cite{PS05} to create
hidden bits from secret help, which in turn uses ideas from \cite{Kil88} in order to do Oblivious Transfer.
For clarity of exposition, we show how to hide a single bit, but the construction naturally generalizes to $n$ bits by repeating in parallel.

\begin{proposition}\label{sha=hb}
Let three random bits $(s^0,s^1,b)$ be such that: the prover knows $s^0$ and $s^1$ and has no information about $b$; the verifier knows $b$ and the associated bit $s^b$ but has no information about $s^{\overline{b}}$. Then we can create a hidden bit $r$ with security parameter $k = 1$.
\end{proposition}
\begin{proof}
From these bits, the associated hidden bit will be $r = s^0 \oplus s^1$ and $s^0, s^1$ will be called the shares of $r$. The way the prover will reveal $r$ is by sending these two shares to the verifier who checks that they correspond with the one share he has. We now show that $r$ is a hidden bit with security parameter $1$:
\begin{itemize}
\item Since $s^0$ and $s^1$ are random and known to the prover, then so is $r$.
\item Since the verifier knows $s^b$ but has no information about $s^{\overline{b}}$, he has no information about $r$.
\item \modif{If the prover tries to lie about $r$ then he has to flip exaclty one of the two shares. He will get caught if he flips $s^b$ and will not get caught if he flips $s^{\overline{b}}$. Since he has no information about $b$, he will be caught cheating with probability $1/2$.}
\end{itemize}
\end{proof}

Note that if we have for each hidden bit $r$, $k$ independent random couples of shares $(s^0_k,s^1_k)$ such that $s^0_k \oplus s^1_k = r$
then similarly, we can suppose that $r$ is a hidden-bit with security parameter $k$.

\subsubsection{A quantum way of achieving Hidden Bits}

From the coins of the verifier, we now show how to create the shares described in the previous part. As before, we describe the construction of one hidden-bit which easily generalizes to $n$ bits.
We use three qubits of the verifier's initial \modif{totally} mixed state (three coins) as
$\sum_{b,s^{b},c \in \01} \ket{b,s^{b},c}\bra{ b,s^{b},c}$.

As in the previous part, the bit $b$ corresponds to which share the Verifier has and $s^b$ corresponds to the value of that share. The bit $c$ corresponds to the value of the other share in the Hadamard basis, i.e. we define $\ket{c^\times} = \frac{1}{\sqrt{2}}(\ket{0}+(-1)^c\ket{1})$. The verifier performs the unitary $U_{b,s^{b},c}$ that depends on $(b,s^b,c)$ and sends the outcome to the Prover.
\[
U_{0,s^{b},c}: \ket{0}\ket{0} \rightarrow \ket{s^{b}} \ket{c^\times} \;\;\; \mbox{ and } \;\;\;
U_{1,s^{b},c}: \ket{0}\ket{0} \rightarrow \ket{c^\times} \ket{s^{b}}
\]

The prover has two qubits which he measures in the computational basis and the outcomes of this measurement will correspond to the two shares. One of this measurements will give $s^b$ and the other one will give a random bit $s^{\overline{^b}}$. The hidden bit $r$ is equal to $s^b \oplus s^{\overline{^b}}$. 

\begin{lemma}\label{c=sha}
The above construction results in a hidden bit $r$ with security parameter~$1$.
\end{lemma} \newpage
\begin{proof}
\begin{itemize}
\item The bit $r = s^b \oplus s^{\overline{^b}}$ is random since the verifier picks $s^b$ at random and the outcome of the measurement of $\ket{c^\times}$ in the computational basis is also random (hence $s^{\overline{b}}$ is random). Since the prover knows the two shares he knows $r$.
\item \modif{The verifier knows a share $s^b$ which is random since $b$ is random}. He has no information about the \modif{share} $s^{\overline{b}}$ since the outcome of the Prover's measurement of $\ket{c^\times}$ is independent of the Verifier's coins. Hence he has no information about $r$.
\item $b$ is unknown to the prover: to show this, let $\rho_b$ be the state of the prover conditioned on the verifier's coin $b$
\[
\rho_0 = \left\{
\begin{array}{l}
wp. \ 1/4 \quad \ket{0,+} \\
wp. \ 1/4 \quad \ket{0,-} \\
wp. \ 1/4 \quad \ket{1,+} \\
wp. \ 1/4 \quad \modif{\ket{1,-}} \\
\end{array}
\right. \;\;\;\;\; \mbox{and} \;\;\;\;\;
\rho_1 = \left\{
\begin{array}{l}
wp. \ 1/4 \quad \ket{+,0} \\
wp. \ 1/4 \quad \ket{+,1} \\
wp. \ 1/4 \quad \ket{-,0} \\
wp. \ 1/4 \quad \ket{-,1} \\
\end{array}
\right.\]
We can easily see that $\rho_0 = \rho_1$ hence the prover has no information about $b$. \modif{Moreover, since $\rho_0 = \rho_1 = \mathbb{I}$, the prover's state is equivalent to a mixture of classical pairs of shares. Since he has no information about $b$, the prover cannot cheat for any of those classical pairs of shares with probability striclty greater than $1/2$}.
\end{itemize}
\end{proof}

We can easily extend the above construction to a hidden-bit with security parameter $k$ for any polynomial $k$ (by creating $k$ independent pairs of shares for this hidden-bit) and also to $n$ hidden-bits with security parameter $k$ by just repeating this process $n$ times.

Note also that the unitary used by the verifier uses his coins only as control bits. Therefore, we can use this construction to create hidden bits in a way which is consistent with our enhanced notion of simulation and show that $HVSZK^{HB} \subseteq HVQSZK^C$. Let us  prove this fact formally:

\begin{proposition}
$HVSZK^{HB} \subseteq HVQSZK^C$ \end{proposition}
\begin{proof}

Let $\Pi$ a problem in $HVSZK^{HB}$ and $\prot{P,V}$ a classical zero-knowledge protocol with hidden-bits that solves $\Pi$.
We create the following quantum protocol $\langle P',V' \rangle$ where the verifier starts with the state : $(\ket{0}\bra{0})^{\otimes n} \otimes \mathbb{I}_n$ (acting as his workspace and coins).
\begin{itemize}
\item The verifier $V'$ views his coins as the coins of the original verifier $V$ and the coins needed in order to create hidden bits.
\item In the beginning of the protocol the verifier uses our construction and creates hidden bits with security parameter $k$.
\item Then, the prover and verifier both follow the original classical protocol $\prot{P,V}$. \modif{Note that this is possible since any classical circuit $C$ can be transformed into a quantum unitary circuit $U_C$ such that $U_C(\ket{x,0}) = \ket{x,C(x)}$.} 
\end{itemize}

Note that since $V'$ uses his coins as the private randomness of $V$, he can perform the classical protocol $\prot{P,V}$ using unitaries.

We now prove that $\langle P',V' \rangle$ is a Zero-Knowledge protocol
that solves $\Pi$.
Completeness is straightforward from the completeness of the original protocol and the fact that in our construction the prover can always reveal the correct hidden bits. Concerning soundness:
\begin{enumerate}
\item If the prover reveals all the hidden-bits correctly, the soundness of $\langle P',V' \rangle$ is the same as the soundness of $\prot{P,V}$.
\item If the prover lies on at least one of the hidden-bits he reveals, then the soundness of $\prot{P',V'}$ will be smaller than $2^{-k}$ since the hidden-bits created have security parameter $k$.
\end{enumerate}

To show the zero-knowledge property, we use the fact that we can already simulate the verifier's view in the protocol $\prot{P,V}$. This includes the private coins of $V$, the messages and in particular, all the hidden-bits $r_i$ revealed by the prover.

In order to simulate the verifier's view in the new protocol $\prot{P,V}$ we have to additionally simulate the following:
\begin{itemize}
\item all the coins that the verifier $V'$ used in order to create hidden bits.
\item The $k$ pairs of shares $(s_i^0,s_i^1)_{j}, j \in [k]$ for every revealed hidden bit $r_i$.
\end{itemize}

First, the simulator just flips some coins in order to simulate all the random bits the verifier uses to construct the hidden bits. In particular, for every revealed hidden bit $r_i$, the simulator has the corresponding bits $(b_i,s^b_i,c_i)_{j}, j \in [k]$.
From these bits and the value of $r_i$ (which we know from the original simulation), we can now create all the couples of shares $(s_i^0,s_i^1)_j$.
This allows us to simulate the view of the verifier in the protocol $\prot{P',V'}$.
\end{proof}

\begin{theorem} \label{coin}
$PSPACE \subseteq HVQSZK^{C}$
\end{theorem}

\begin{proof}
From Section \ref{hidden-bits} we know that: $PSPACE \subseteq HVSZK^{HB}$. We now use the fact that $HVSZK^{HB} \subseteq HVQSZK^{C}$ and conclude.
\end{proof}

One might think that this surprising result comes from the fact that the round-by-round simulation is too weak in our setting and that a satisfactory zero-knowledge property is not achieved. In fact, if we assume that the verifier follows the protocol, then our notion of simulation is as strong as in the unitary case. The only extra information that the verifier has in our protocols is the initial random string which we always simulate at every round.

\subsection{Forcing the honest behaviour of the verifier}

From a cryptographic point of view, the important question is when we can actually assume that the verifier behaves in an honest way during a protocol.
Watrous' result that $QSZK = HVQSZK$ shows that if the protocol only asks the verifier to perform unitary operations, then we can actually force the verifier to behave honestly. 
Our result shows that this is probably very difficult to achieve unconditionally when the protocol asks the verifier to additionally flip coins, since in that case $IP \subseteq QSZK$. Note that this is a striking difference with the classical case where the verifier can perform any classical operation (including coin flips). 

However, in certain realistic settings, we can assume that the verifier is semi-honest, for example when he interacts with the prover via a secure interface like an ATM or a secure web page. We are going to define a classical and a quantum model for this type of interaction in the following way.
In the classical model, the prover and the verifier interact via a $\emph{deterministic}$ machine whose behaviour is known both to the prover and the verifier. The verifier interacts with the prover by providing a classical input to the machine which after performing some computation on this input, sends a message to the prover. On the other hand, the machine transmits the messages of the prover to the verifier unchanged. It is easy to see, that the languages that have a statistical zero knowledge protocol in this model are exactly the ones in the class of Honest Verifier Statistical Zero Knowledge ($HVSZK$), since an honest verifier can always act as the deterministic machine and vice versa. Since $HVSZK = SZK$, we conclude that this type of interface does not increase the possibility to do zero-knowledge protocols.

We can define a quantum analog of this model in the following way. 
The prover and the verifier interact via a $\emph{deterministic}$ machine whose behaviour is known both to the prover and the verifier. The verifier interacts with the prover by providing a {\em classical} input to the machine which after performing some computation on this input, creates a {\em pure quantum} state and shares it between the prover and verifier. On the other hand, the machine transmits the messages of the prover, which could be mixed quantum states, to the verifier unchanged. 
First, we can see that this model contains the class of Honest Verifier Quantum Statistical Zero Knowledge even when we allow the verifier to flip coins, since for any protocol there exists a machine that takes as input the private coins of the verifier and performs the same unitary operations as the verifier.
Hence, unlike the classical case where this semi-honest model is no more powerful than $SZK$, in the quantum semi-honest case, we have zero-knowledge protocols for any problem in $PSPACE$ (Section \ref{HVQSZK}). 

Let us also note, that if we allow the verifier to provide quantum input to the machine, then the above model is exactly the class $QSZK$. It would be very interesting to see what are the most general zero knowledge protocols for which  we can force a quantum verifier to behave honestly.

%%%%%%%%%%%%%%%%%%%%%%%%%%%%%%%%%%%%%%%%%%%%%%%%%%%%%%%%%%%%%%%%5
%%%%%%%%%%%%%%%%%%%%%%%%%%%%%%%%%%%%%%%%%%%%%%%%%%%%%%%%%%%%%%%%5
%%%%%%%%%%%%%%%%%%%%%%%%%%%%%%%%%%%%%%%%%%%%%%%%%%%%%%%%%%%%%%%%5
%%%%%%%%%%%%%%%%%%%%%%%%%%%%%%%%%%%%%%%%%%%%%%%%%%%%%%%%%%%%%%%%5
%%%%%%%%%%%%%%%%%%%%%%%%%%%%%%%%%%%%%%%%%%%%%%%%%%%%%%%%%%%%%%%%5

\section{Non-unitaries and cheating verifiers} \label{QSZK_NU}
\subsection{Definitions}
The goal of this section is to describe Watrous' definition of Quantum Statistical Zero Knowledge ($QSZK$) for cheating verifiers.
Consider a quantum zero-knowledge protocol between a prover $P$ and a verifier $V$ where the verifier starts with an auxiliary input $w$. Additionally, the prover and verifier have as common input the input of the promise problem which is a classical string. All the operations described hereafter will depend on this input and this dependence will be omitted.

We will use the following Hilbert spaces for our analysis.
\begin{itemize}
\item $\spa{P}$ the space of the prover.
\item $\spa{M}$ the space where the prover and verifier store the messages they send.
\item $\spa{V}$ the verifier's workspace initialized to $\ket{0}$.
\item $\spa{W}$ the verifier's space where the auxiliary input is initially stored.
\end{itemize}

Let $\langle{P,V}\rangle = \langle{P_1,V_1,\dots,P_n,V_n}\rangle$. Each $P_i$ acts on $\mathcal{P} \otimes \mathcal{M}$ and each $V_i$ acts on $\mathcal{M} \otimes \mathcal{V} \otimes \mathcal{W}$. We can tensor these operations with the identity and suppose that they all act on the space: $\spa{P} \otimes \spa{M} \otimes \spa{V} \otimes {W}$. We can therefore see the whole protocol as a big operation $O$ acting on $\spa{P} \otimes \spa{M} \otimes \spa{V} \otimes {W}$. More formally:

\begin{definition}
For any protocol $\langle{P_1,V_1,\dots,P_n,V_n}\rangle$ where each $V_i$ and $P_i$ acts on $\spa{P} \otimes \spa{M} \otimes \spa{V} \otimes {W}$ (in fact by tensoring the $V_i$'s and $P_i$'s with the identity) we denote by $O_{P,V}$ the following admissible mapping:
\[
\begin{array}{lll}
O_{P,V} &: \spa{L}(\spa{W}) &\rightarrow \spa{L}(\spa{P} \otimes \spa{M} \otimes \spa{V} \otimes {W}) \\
&: w &\rightarrow V_n(P_n(\dots(V_1(P_1(\underbrace{\ket{0}}_{\in\spa{P}\otimes\spa{M}\otimes\spa{V}} \otimes \underbrace{w}_{\in \mathcal{W}})))))
\end{array}
\]
\end{definition}
where $\spa{L}(\spa{X},\spa{Y})$ is the set of linear operators from $\spa{X}$ to $\spa{Y}$, and $\spa{L}(\spa{X}) = \spa{L}(\spa{X},\spa{X})$. In particular, any mixed state in $\spa{X}$ can be represented as an element of $\spa{L}(\spa{X})$. \\

The zero-knowledge property concerns only what the verifier has at the end of the protocol. Without loss of generality, we can suppose that $\spa{M}$ is empty since a cheating verifier can always move the information from $\spa{M}$ to $\spa{V}$ at the end of the protocol. Hence, we will be interested in:
\[
\begin{array}{lll}
O_{V} &: \spa{L}(\spa{W}) &\rightarrow \spa{L}(\spa{V} \otimes {W}) \\
&: w &\rightarrow Tr_{\spa{P} \otimes \spa{M}}\left( V_n(P_n(\dots(V_1(P_1(\underbrace{\ket{0}}_{\in\spa{P}\otimes\spa{M}\otimes\spa{V}} \otimes \underbrace{w}_{\in \mathcal{W}})))))\right)
\end{array}
\]
which for short we will also denote as $O_V = Tr_{\spa{P} \otimes \spa{M}}O_{P,V}$. \modif{More generally, for any super-operator X that outputs in $\spa{A} \otimes \spa{B}$, we denote $Tr_\spa{A} X$ the super-operator such that $(Tr_\spa{A} X)(\rho) = Tr_\spa{A}(X(\rho))$.}
We say that $O_V$ is the mapping that corresponds to the verifier's view of the protocol. We want to be able to simulate this mapping $i.e.$ be able to create in quantum polynomial time a mapping $\Sigma$ which will act like $O_V$ and this for every auxiliary input $w$. We can now define $QSZK$:
\begin{definition}
We say that $\Pi \in QSZK$ if there is a protocol $\langle P,V \rangle = \langle{P_1,V_1,\dots,P_n,V_n}\rangle$ such that:
\begin{itemize}
\item Completeness: $\forall x \in \Pi_Y$, the verifier accepts with probability greater than $2/3$.
\item Soundness: $\forall x \in \Pi_N$, and for all prover's strategies $P^*$, the verifier accepts with probability smaller than $1/3$.
\item Zero-knowledge: for any cheating verifier $V^*$ (where $O_{V^*}$ is the mapping associated to $\langle P,V^* \rangle$), there is a function $\mu$ and a mapping $\Sigma : \spa{L}(\spa{W}) \rightarrow \spa{L}(\spa{V} \otimes {W})$ that can be computed in quantum polynomial time such that $\forall x \in \Pi_Y$, we have
\[
||O_{V^*} - \Sigma||_{\diamond} \le \mu(|x|). \]
\end{itemize}
\end{definition}
where for any super-operator $\Phi$, $||\Phi||_{\diamond} = sup\{||\Phi \otimes I_{\spa{L}(\spa{Z})}||_{tr}, \ \spa{Z}$ is a complex Euclidean space$\}$ (see~\cite{KSV02} for more details on this diamond norm).

Note that if $\Sigma$ uses $V^*$ only as a black box, then we can change the order of quantifiers and have a single mapping $\Sigma$ for all possible $V^*$.

In the definition of $QSZK$, the verifier and the prover can use any physically admissible operation. We will show that in fact, if the zero-knowledge property holds against cheating verifiers that only use unitaries then it also holds for cheating verifiers that use any physically admissible operation. In other words, cheating strategies with unitary operations are the most general ones.

\begin{definition}
We say that $\Pi \in QSZK^U$ if there is a protocol $\langle P,V \rangle = \langle{P_1,V_1,\dots,P_n,V_n}\rangle$ such that:
\begin{itemize}
\item Completeness: $\forall x \in \Pi_Y$, the verifier accepts with probability greater than $2/3$.
\item Soundness: $\forall x \in \Pi_N$, and for all prover's strategies $P^*$, the verifier accepts with probability smaller than $1/3$.
\item Zero-knowledge: for any cheating verifier $V^*$ that uses unitaries (where $O_{V^*}$ is the mapping associated to $\langle P,V^* \rangle$), there is a function $\mu$ and a mapping $\Sigma : \spa{L}(\spa{W}) \rightarrow \spa{L}(\spa{V} \otimes {W})$ that can be computed in quantum polynomial time such that $\forall x \in \Pi_Y$, we have
\[
||O_{V^*} - \Sigma||_{\diamond} \le \mu(|x|). \]
\end{itemize}
\end{definition}

\subsection{Unitary cheating verifiers are as powerful as general cheating verifiers}

In this section, we show that in the case of cheating verifiers, coin flips $-$ and more generally any non-unitary operations $-$ do not add anything to the power of quantum Zero-Knowledge. In other words, we show that

\begin{proposition}
$QSZK = QSZK^{U}$
\end{proposition}

\begin{proof}
We have by definition that $QSZK \subseteq QSZK^U$. We show now the other inclusion. The main idea is to say that each time the verifier uses a non-unitary, he can use a larger unitary which will act as a purification of this non-unitary which will only give him more information. More formally, we use the following fact that is a direct corollary of the purification lemma. (see \cite{NC00}).
\begin{lemma}\label{nucirc} Let $C$ a quantum non-unitary circuit acting on a space
$A$. There is a space $\spa{B}$ of same dimension as $\spa{A}$ and a unitary circuit $\widetilde{C}$ acting on $\spa{A} \otimes
\spa{B}$ such that $Tr_{\spa{B}}\widetilde{C} = C$.
\end{lemma}
Now consider a protocol $\langle P,V \rangle = \langle{P_1,V_1,\dots,P_n,V_n}\rangle$ which has the zero-knowledge property for any unitary cheating verifier $V$. Consider a cheating verifier $V^*$, the protocol $\langle P,V^* \rangle = \langle{P_1,V^*_1,\dots,P_n,V^*_n}\rangle$ and its associated mapping $O_{V^*}$ from $\spa{L}(\spa{W})$ to $\spa{L}(\spa{V} \otimes \spa{W})$. Recall that:
\[
O_{V^*} = Tr_{\spa{P} \otimes \spa{M}} \left( P_n \circ V^*_n \circ \dots \circ P_1 \circ V^*_1 \right) \]
Consider now $n$ additional Hilbert spaces $\spa{A}_1$ through $\spa{A}_n$ and admissible mappings $\widetilde{V}^*_i$ such that
\[
\forall i \ Tr_{\spa{A}_i}\widetilde{V}^*_i = V^*_i \]
The spaces $\spa{A}_i$ are Hilbert spaces that the verifier possesses.
Let us  look at the protocol $\langle P,\widetilde{V}^* \rangle = \langle{P_1,\widetilde{V}^*_1,\dots,P_n,\widetilde{V}^*_n}\rangle$
and $O_{\widetilde{V}^*}$ the associated mapping for the verifier. This mapping is a mapping from  $\spa{L}(\spa{W})$ to $\spa{L}(\spa{A}_1 \otimes \dots \otimes \spa{A}_n \otimes \spa{V} \otimes \spa{W})$. We know that there is a mapping $\Sigma$ computable in quantum polynomial time such that
$
||O_{\widetilde{V}^*} - \Sigma||_{\diamond} \le \mu(|x|). $

By construction, we know that $O_{V^*} = Tr_{\spa{A}_1 \otimes \dots \otimes \spa{A}_n}O_{\widetilde{V}^*}$. Consider $\Sigma' = Tr_{\spa{A}_1 \otimes \dots \otimes \spa{A}_n}\Sigma$, we can easily conclude that
\[
||O_{V^*} - \Sigma'||_{\diamond} \le \mu(|x|) \]
and that $\Sigma'$ is quantum polynomial time computable which concludes our proof.
\end{proof}

\section{Conclusion and further work}

We showed that the unitarity restriction in quantum Zero Knowledge is not inconsequential. In the case of Honest-Verifier, we showed that allowing the verifier to flip coins is sufficient to construct quantum Statistical Zero Knowledge protocols for any Interactive Proof. This is the first time that Statistical Zero Knowledge was achieved for such a large class unconditionally, even in the case of honest verifier. We believe that it is a strong witness of the fact that $HVQSZK^{C} \neq HVQSZK$ and therefore that coin flips increase substantially the power of Honest-Verifier Quantum Zero Knowledge. We also showed that this difference does not hold when dealing with cheating verifiers.

A first question concerning our result is whether it is possible to use it in realistic zero-knowledge protocols. While it seems improbable that our protocols can be transformed unconditionally into a protocol secure against cheating verifiers, this may be done using a 3rd party or computational assumptions. As future work, it would be interesting to develop a more realistic model for a quantum semi-honest adversary. 

Moreover, protocols which satisfy some weaker zero-knowledge properties are used in other cryptographic applications and it is also interesting to consider what can be done in the quantum setting. Finally, the main question that remains open is whether it is in fact possible to achieve cheating-verifier quantum Statistical Zero Knowledge for some classical language not in $SZK$ or $BQP$.

\newcommand{\etalchar}[1]{$^{#1}$}

%\newpage

\appendix 

\section{Solving a $QMA$-complete problem in $HVQSZK^{C}$}\label{QMA}

We show here how a $QMA$-complete problem can be solved using a protocol with Hidden-Bits. Note that we already know that all of $QMA$ can be done using hidden-bits. However, the protocol we will show has constant rounds and relies in fact only on the ability to do bit commitment.
We transform a $QMA$ protocol for the $QMA$-complete problem $LCDM$ defined by Liu~\cite{Liu06} into a Zero Knowledge one.  \\ \\
{\em Local Consistency of Density Matrices} $(LCDM)$\\
\textbf{Input :} a number $n$, a set $L \subseteq \{1,\dots,n\} \times
\{1,\dots,n\}$ and matrices $M_{i,j} \textrm{ of size } 4\times 4$
for all $(i,j) \in L$. These matrices have $t = poly(n)$ bits of precision. \\
\textbf{Promise }

\underline{Yes instance}: There exists a quantum pure state $\ket{\phi}$ such that $\forall (i,j) \in L$, the reduced density matrix of $\ket{\phi}$ on the pair of qubits $(i,j)$ is equal to $M_{i,j}$.

\underline{No instance}: For any quantum pure state $\ket{\phi}$, there is a couple $(i,j) \in L$, such that the reduced density matrix of $\ket{\phi}$ on the pair of qubits $(i,j)$ has trace distance greater than $1/t$ from $M_{i,j}$.

\paragraph{$QMA$ protocol for $LCDM$:}
Let $x$ an instance of $LCDM$.
The verifier receives a state $(w_{1},\dots,w_{K})$ that corresponds (if $x \in LCDM_Y$) to $K$ copies of a witness of $x$. He then picks
$(l_{1},\dots,l_{K}) \in L$ at random. For every witness ${w_i}$ he traces out all but the qubits in $l_i$ and hence has
\modif{$w_{i}^{l_{i}} = Tr_{\{1,\dots,n\} \times \{1,\dots,n\} - l_i}w_{i}$}. He then applies an accepting procedure
$A(w_{1}^{l_{1}},\dots,w_{K}^{l_{K}})$.\\ \\
Liu showed the following:
\begin{proposition}
There exists  $K \in poly(n)$ and an accepting procedure $A$ for the verifier such that the above procedure has completeness $1 - 2^{-n}$ and soundness $2^{-n}$.
\end{proposition}
In particular this shows that if $x \in LCDM_N$ the prover cannot cheat, even by sending an entangled state.
We describe now a quantum statistical zero-knowledge protocol that will solve $LCDM$ using hidden-bits.\\
\cadre{
\begin{center}
\textbf{ Protocol in $HVQSZK^{HB}$ for the $LCDM$ problem }
\end{center}
{\bf Input:} An instance $\Pi$ of the $LCDM$ problem \\ \\
\textbf{DO} K times in parallel : \\
\textbf{HB} : Use as a random string $r = (r_{1},s_{1},r_{2},s_{2},\dots,r_{n},s_{n})$ \\
\textbf{P} : Let $\ket{w}$ be the witness in $\Pi$. Send $U_r \ket{w}$ to the Verifier.\\
%Create $\eta = U(|\tilde{r}\rangle,|\omega\rangle) = |\tilde{r}\rangle %U_{\tilde{r}}|\omega\rangle$  \\
%send the second half $B$ of $\eta$ to the verifier. \\
\textbf{V} : Pick $l =(x,y) \in_{R} L$  and send $l$ to the prover. \\
\textbf{P} : Reveal $(r_{x},s_{x})$ and $(r_{y},s_{y})$  to the verifier. \\
\textbf{V} : Apply $\left( X^{r_{x}}Z^{s_{x}} \right)^\dagger$ (resp. $\left( X^{r_{y}}Z^{s_{y}} \right)^\dagger$) to the $x^{th}$ (resp. $y^{th}$) qubit; trace out the others. \\
\textbf{END} \\ 
Let $z^{1},\dots,z^{K}$ the density matrices of the qubits the Verifier kept in each repetition \\
\textbf{V} : Compute $A(z^{1},\dots,z^{K})$.}

Let $r = (r_1,s_1,\ldots, r_n,s_n)$ a $2n$-bit string and $U_r$ a unitary that acts independently on $n$ qubits such that  $U_r = X^{r_1} Z^{s_1} \otimes \ldots \otimes X^{r_n} Z^{s_n}$, where $X,Z$ are the bit- and phase-flip operators. $U_r$ performs a perfect encryption of an $n$-qubit state, i.e for any $n$-qubit state $\rho$ we have \modif{$\frac{1}{2^{2n}} \sum_r U_r \rho U_r^\dagger  = \mathbb{I}$}. \\

%\paragraph{Proof of the protocol}
\noindent
{\bf Completeness and soundness} For the $K$ parallel
repetition, the prover chooses $K$ witnesses $w_{1},\dots,w_{K}$ similarly than in the $QMA$-protocol for $\Pi$ and the verifier's states at the end are $ z^{1},\dots,z^{K}$.
Since the prover cannot lie on his hidden-bits and the
verifier chooses the $l$'s at random, the verifier has at the end
exactly $w_{1}^{l_{1}},\dots,w_{K}^{l_{K}}$ with the
$l_{K}$ chosen at random. Completeness and soundness follows from Liu's analysis.

\noindent
{\bf Zero Knowledge}
We need to simulate the Verifier's view after the Prover's first message ($\rho_1$) and after the second message ($\rho_2$).
The state $\rho_{1}$ consists of the message sent by the prover. From the
analysis of $U$, we know that the verifier's view of this state is
the totally mixed state. This is because he has no information about the hidden bits $r$. Therefore, $\rho_{1} = \mathbb{I}$ and can be easily simulated.

The state $\rho_{2}$ is trickier. The verifier has a state $\rho$ that is the totally mixed state, a random $l =(i,j) \in L$ and bits $(r_i,s_i,r_j,s_j)$ such that he can decode the qubits $(i,j)$ of $\rho$ and transform his state into one which is totally mixed except on the qubits $(i,j)$, where the density matrix is $M_{i,j}$.

We simulate this as follows: we pick $l \in_R L$ and some state $\ket{\alpha_l}$, such that the reduced density matrix on the qubits in $l=(i,j)$ is $M_{i,j}$. Then we pick a random string $r = (r_1,s_1,\ldots,r_n,s_n)$ and apply $U_r$ to the state $\ket{\alpha}$. More precisely, we first create the quantum state
$ \sum_{l,r} \ket{l}\ket{r}\ket{l}U_r(\ket{\alpha_l})\ket{r_i,s_i,r_j,s_j}
$
and then trace out the first two registers. The resulting state simulates the view of the Verifier after the Prover's second message. We conclude that this problem is in $HVQSZK^{HB}$ and hence in  $HVQSZK^{C}$.

Note that the $LCDM$ problem is $QMA$-complete via randomized Turing reductions and hence this protocol cannot be used for all languages in $QMA$. Showing that $LCDM$ is $QMA$-complete via mapping reductions is still an open problem.

\end{document}